# A Scalable AI Driven, IoT Integrated Cognitive Digital Twin for Multi-Modal Neuro-Oncological Prognostics and Tumor Kinetics Prediction using Enhanced Vision Transformer and XAI

**Himadri Nath Saha[1] (Senior Member, IEEE), Utsho Banerjee[1], Rajarshi Karmakar[2], Saptarshi Banerjee[3], Jon Turdiev[4]**

[1]Department of Computer Science, SNEC, University of Calcutta, Kolkata, India (e-mail: contactathimadri@gmail.com)
[1]Bachelor of Computer Science and Engineering(Specialization in Internet Of Things), Institute of Engineering and Management, West Bengal, India (e-mail:utshobanerjee1602@gmail.com)
[2]Bachelor of Computer Science and Engineering, University of Engineering and Management, West Bengal, India (e-mail: karmakar1raj@gmail.com)
[3]Illinios Institute of Technology, Chicago, USA (e-mail: banerjee.saptarshi44@gmail.com)
[4]San Francisco State University, San Francisco, USA(e-mail: nturdiev@gmail.com)

Corresponding author: Himadri Nath Saha (e-mail: contactathimadri@gmail.com)

**ABSTRACT** Neuro-oncological prognostics are now vital in modern clinical neuroscience because brain tumors pose significant challenges in detection and management. To tackle this issue, we propose a cognitive digital twin framework that combines real-time EEG signals from a wearable skullcap with structural MRI data for dynamic and personalized tumor monitoring. At the heart of this framework is an Enhanced Vision Transformer (ViT++) that includes innovative components like Patch-Level Attention Regularization (PLAR) and an Adaptive Threshold Mechanism to improve tumor localization and understanding. A Bidirectional LSTM-based neural classifier analyzes EEG patterns over time to classify brain states such as seizure, interictal, and healthy. Grad-CAM-based heatmaps and a three.js-powered 3D visualization module provide interactive anatomical insights. Furthermore, a tumor kinetics engine predicts volumetric growth by looking at changes in MRI trends and anomalies from EEG data. With impressive accuracy metrics of 94.6% precision, 93.2% recall, and a Dice score of 0.91, this framework sets a new standard for real-time, interpretable neurodiagnostics. It paves the way for future advancements in intelligent brain health monitoring.

**INDEX TERMS** Neuro-oncological prognostics, Digital Twin, ViT++ , Wearable Skullcap, EEG, MRI, Tumor Kinetics.

## I. INTRODUCTION

Recent advancements in brain-computer interface (BCI) systems, real-time neuroimaging, and edge-native computation have opened a new frontier in intelligent neurological diagnostics. A key part of this technological shift is the digital twin, which is a virtual model that updates in real-time and represents the structural and functional traits of its physical counterpart. While digital twins have shown significant promise in areas like aerospace, industrial automation, and personalized medicine, their use in cognitive-neurophysiological modeling is still lacking and not fully developed.Current frameworks that try to replicate brain dynamics through digital twins have several key limitations. Many of these models are fixed in their structure and rely on past data without continuously incorporating real-time physiological signals. There are those restricted to one of the two kinds of data: either anatomical neuroimaging (such as MRI) or electrophysiological recordings (such as EEG). They hardly integrate these to form a complete neurocognitive image. Many of them greatly depend on the use of cloud-based systems that may lead to delays, can not be used in regions with bad connections, and impose some serious concerns related to data safety and patient confidentiality. Above all, these systems rely in principle on non-transparent and complex models of AI. The problem is that such indistinctness impacts the interpretation of the results not only by clinicians but also erodes trust in the diagnosis. This is important as far as making key decisions regarding neurological health is concerned. Some of these challenges have been attempted to be addressed by the recent research. Specifically, Zhihan Lv et al. [2] suggested a BCI-centric digital twin that applies advanced Riemannian manifold-based transfer learning models to a better EEG-based motor intention classification. Although this study enhanced the interpretation of functional signals, it was restricted to the EEG and did not incorporate structural imaging, real-time flexibility, and clinical explainability. On the same note, a multimodal image fusion strategy driven by deep transfer learning jointly used MRI and PET/SPECT imaging to enhance diagnostic accuracy was investigated by Jinxia Wang et al. [3]. Despite some advancement of spatial detail survival and modal synergy, the system was largely off-line, not connected to physiological signals, and could not sustain adaptive change in time; all of these are essential elements of a realistic model of cerebral behaviour. All these drawbacks emphasize the necessity of a holistic, real-time and edge-compatible digital twin architecture that would meet the current needs of neuroscience. The described framework needs to have an ability to complement functional and structural neurodata and be free to operate at the edge without the need of external maintenance, adjust dynamically to the new inputs and not lose a readership in its rationale. It must not only be a model of static diagnosis, but a cognitive-neurophysiological substitute that could reflect, analyze and simulate a patient's cerebral state in a clinically actionable way.

We suggest a consistent, scalable, intelligent digital twin architecture of continuous brain health monitoring and tumor progression analysis. New contributions to this work are a wearable skull cap with a custom EEG interface, edge computing, data authentication and confidence analysis using an adaptive risk filtering mechanism that avoids redundant cloud communication and a hybrid digital twin in which both Vision Transformer-based MRI segmentation and EEG-based brain state detection are combined. The digital twin framework also uses explainable AI techniques, such as Grad-CAM, for clear tumor localization, features a 3D interactive brain model for detailed risk and tumor visualization, and includes a tumor kinetics engine that simulates how tumors grow over time and space, helping with real-time monitoring and treatment planning. This interpretable solution addresses the flaws of earlier disconnected and unclear systems, providing a practical and personalized setup for continuous brain health monitoring and proactive neuro-oncology management.

The paper is organized as follows: Section 1 introduces the motivation, problem statement, objectives, and our new contributions; Section 2 reviews the current state-of-the-art architectures in neuro-medical diagnosis; Section 3 offers a detailed description of the dataset; Section 4 and Section 5 outline the system architecture and methodology; Section 6 discusses the experimental results; and Section 7 wraps up with key contributions and future research directions.

## II. RELATED WORKS

Digital twin technology has become a game-changer for simulating and monitoring neurophysiological conditions using real-time data, AI models, and multiple data types. Despite progress in deep learning-based diagnostics, signal decoding, and immersive analysis, most current systems struggle with scalability, real-time responsiveness, explainability, and combining both structural and functional brain data. To tackle these issues, Aftab Hussain et al. [1] proposed an attention-based ResNet-152V2 model for detecting and classifying intracranial hemorrhage (ICH), contributing to Health 4.0 digital twin applications. This architecture involves the process of extracting features of interest (attention mechanism), dimensionally reducing them (principal component analysis (PCA)), and creating data with a less prevalent subtype of ICH as an input layer with the help of a deep convolutional generative adversarial network (DCGAN). On the RSNA 2019 dataset, the model demonstrated great accuracy in different hemorrhage types, with values above 99 for epidural hemorrage and above 97 when it comes to intraparenchymal hemorrage. Nevertheless, despite its effectiveness and high classification accuracy, the fact that the model relies on synthetic data is suspect to overfitting and bias. In addition, the study fails to indicate its ability to be generalized to other clinical datasets or practical scenarios.

Future research, as proposed by the authors, should be aimed at testing larger datasets and incorporating explainability, considering that the model in question currently lacks transparency, which is a fundamental limitation when using it in a clinical setting in the Health 4.0 context. Developing the neurofunctional part, Zhihan Lv et al. [2] proposed a cognitive computing model of brain-computer interface (BCI)-based digital twins, with the intention of interpreting electroencephalography (EEG) signals. They used different preprocessing and feature extraction methods of EEG, which encompassed Butterworth and FIR filters, wavelet decomposition, and a new TL-TSS algorithm involving the Riemannian manifold theory. To decode the EEG signal they employed a hybrid entropy and singular spectrum analysis (SSA) approach. TL-TSS approach exceeded other conventional methods such as Common Spatial Pattern (CSP), realising a classification accuracies of up to 97.88 percent on the BCI competition datasets. Nevertheless, it can only be used in a motor imagery-type task and is less applicable in other brain disorders like epilepsy, cognitive decline, or neuro-oncology. The paper describes the necessity of the generalization over the users and bigger studies but fails to plan how it can be implemented and integrated with structural imaging in real-time. Also, the system does not use the transformer-based architecture and edge processing, which limits its future scalability. Working on improvement of pictures and maintaining structural integrity. Wang, Jinxia and her teammates [3] created a deep transfer learning framework together with digital twins to enhance the magnetic resonance imaging (MRI) quality and assist with diagnostic decisions. The model has a modified deep neural network that deliberately does not employ batch normalization and employs its own loss function to serve as a convergent factor. A new fusion technique using MRI images with either PET or SPECT images involving the use of adaptive decomposition was also presented in the research that preserved spatial information and clinical detail. The quantitative assessment revealed a maximum value of signal-to-noise ratio (PSNR) of 34.11 dB and structural similarity index measure (SSIM) of 85.24% points at its superiority over the current solutions. Nevertheless, this paradigm is greatly dependent on preprocessing, as well as off line analysis, and its real-time usage is poorly investigated. It also does not have closed-loop feedback mechanism of updating dynamic twins and it does not incorporate EEG data nor allow tissue-level visualization and real-time inference that is vital component of continuous neurological monitoring. A visual analytic platform named DTBIA, which enables the user to navigate digital twin simulation through virtual reality, was presented by Yao et al. [4]. This system facilitates multi-resolution connection with the brain signals such as blood-oxygen-level-dependent (BOLD) and diffusion tensor imaging (DTI) signals. This has allowed researchers to be able to interpret what is happening at the voxel level and at the regional level using this 3D visual environment. DTBIA entails hierarchical visualization, 3D edge bundling, and immersive navigation, making it simpler to study the network structure of the brain. Although the system will be beneficial to researchers, it does not serve the practical clinical application because it requires expensive VR devices and graphics processing units (GPUs). Furthermore, the platform is largely exploratory and does not include predictive modeling, input of real-time signals, EEG-based functional analysis. This brings out the necessity towards improved scalability and user accessibility in new versions. To make the system more portable and accessible to use, Sagheer Khan et al. [5] developed RF (radio frequency)-based digital twin of the continuous stroke monitoring utilizing ultra-wideband (UWB) backscatter sensors. They employ machine learning (ML), and deep learning (DL) methods such as stacked autoencoders and fine-tuned k-nearest neighbors (KNN) classifiers. When using the data augmentation strategy of adding Gaussian noise to the data, the model was able to get 93.4 percent and 92.3 percent classification accuracies in binary and multiclass stroke identification, respectively. The portability of this arrangement comes out of its wearable characteristic and real-time feedback. Nonetheless, the model is yet to be applied on real-life hospital or clinical EEG data, which means that the model requires further clinical trials and higher-order signal interpretation algorithms. Also, the digital twin in this case is not proactive but reactive and does not support simulation and forecasting option which is characteristic of today cognitive twin system. The fact that it lacks explainability or the element of 3D visualizations restricts its diagnostic possibility. Upadrista et al. [6] designed a digital twin architecture based on blockchain technology to predict brain stroke. It is a logistic regression based classification system and can use univariate based feature selection with batch gradient descent during training. It plugs synthetic and public data that is stored securely in a consortium blockchain environment that was created using Ganache. This arrangement provides safe information transfer between different healthcare organizations and had a classification score of 98.28 percent, which surpassed the baseline systems. Nevertheless, the architecture is better in terms of data protection but is working with unchangeable sets of data, does not provide data being streamed in real-time, and is not compatible with imaging data types or physiological signals. It cannot be used in dynamic clinical practice due to the absence of the ability to update and visualize. Going to natural human interaction with digital systems. Siyaev et al. [7] proposed a neuro-symbolic reasoning (NSR) framework for voice-based query processing in digital twins. This architecture includes a neural translator based on gated recurrent units (GRUs) that converts spoken input into symbolic logic, which is then processed by a symbolic executor acting on annotated 3D models. Validated on a custom dataset with over 9,000 queries about aircraft maintenance, the system achieved a neuro-symbolic accuracy of 96.2%, a BLEU score of 0.989, and a failure rate of 0.2%. While the model performed well in this context, it is not specific to healthcare, which limits its application for brain modeling. Modifying this architecture for brain digital twins would need the development of specialized symbolic vocabularies and 3D annotated neuroanatomical models, which is both resource-intensive and largely unexplored. The lack of physiological data streams or real-time interaction with multiple inputs further limits the framework's usefulness in neurocognitive applications. Sultanpure et al. [8] introduced a cloud-based digital twin for brain tumor detection, integrating Internet of Things (IoT) devices and deep learning classifiers. The system uses particle swarm optimization (PSO) to select optimal features from MRI scans and evaluates classification performance across convolutional neural networks (CNNs), support vector machines (SVMs), and extreme learning machines (ELMs). CNNs showed the highest tumor detection accuracy. The cloud infrastructure supports centralized data management in line with Healthcare 4.0 standards. However, while CNNs performed exceptionally well on preprocessed MRI scans, the system lacks integrated explainable AI (XAI) techniques such as Grad-CAM and SHAP. The authors also point out potential latency issues because of reliance on cloud services and emphasize the need for real-time feedback loops. Additionally, the current model is not multimodal and does not integrate functional signal analysis like EEG, limiting its ability to provide a comprehensive model of the brain. In a different approach, Wan et al. [9] combined semi-supervised learning with a modified AlexNet to create a digital twin for brain image fusion and classification. This system uses semi-supervised support vector machines (S3VMs) to handle both labeled and unlabeled data, improving model generalization. The enhanced AlexNet speeds up segmentation while also boosting accuracy. The model achieved a recognition accuracy of 92.52%, a Dice Similarity Coefficient of 75.58%, a Jaccard Index of 79.55%, and low error margins, with RMSE and MAE values of 4.91% and 5.59%, respectively. However, the model depends on the manually tuned hyperparameters and is not optimized to use in an online streaming setting. It also does not present functional signal integration, explainability tools and dynamic visualization, and thus restricts the application of the tool in cognitive monitoring. Therefore, more verification in clinical trials is needed to determine the strength of the model and its adaptability adequately. Lastly, Cen et al. [10] adopted a digital-twin modeling to map disease-specific brain atrophy in multiple sclerosis (MS) patients. The authors have investigated the volume of thalamus on the MRI scans to generate an aging curve juxtaposing the MS patients with simulated healthy twins using mixed spline regressions models with different splines (12 types) and different covariate-structure (52). The information was taken via the Human Connectome Project (HCP), the Alzheimer Disease Neuroimaging Initiative (ADNI) and a longitudinal study at one center. The analysis revealed that thalamic atrophy began nearly 5 6 years earlier than the disease was clinically diagnosed showing the earlier biological manifestation of the disease. Cross-validation, Akaike Information Criteria (AIC), and Bayesian Information Criteria (BIC) and bootstrapping were used to reinforce the strength of the model. Nevertheless, the complexity of the model requires large longitudinal data and huge calculations, which is a constraining factor to scalability. It cannot perform real-time update of the data, multi modality fusion and does not support tracking of functional state with evident limitations of further usage and practical implementation. Across the reviewed literature, several common limitations persist. These include a lack of real-time processing, the absence of combining structural (MRI) and functional (EEG) brain signals, limited scalability due to offline or static structures, and minimal use of explainable AI (XAI) techniques like Grad-CAM or SHAP. Furthermore, most systems do not use edge computing, lack dynamic updates, and offer little to no support for 3D visualization, cognitive state tracking, or tumor progression forecasting. These features are crucial for clinical reliability and ongoing neurological monitoring. Our proposed model addresses these limitations through a real-time, multimodal digital twin framework that integrates MRI and EEG data for a thorough brain health analysis. The system has an edge-fog-cloud architecture, where the Raspberry Pi manages EEG preprocessing and the Jetson Nano carries out risk-based filtering. This setup ensures low-latency and portable operation. The cloud-hosted digital twin interface features the enhanced Vision Transformer (ViT++), which provides high-accuracy tumor classification along with XAI visualizations. The model also includes a Tumor Kinetics Engine that predicts tumor growth dynamics over time using patient-specific data. Additionally, the platform allows for interactive 3D brain visualization, real-time cognitive state analysis, and adaptive feedback loops. This offers a scalable, understandable, and clinically useful system that follows the principles of Health 4.0 and next-generation brain healthcare.

| ARCHITECTURE | NOVELTY | EVALUATION METRICS | REFERENCE |
|---|---|---|---|
| Attention-based Residual Network-152V2 (ResNet-152V2) + Principal Component Analysis (PCA) + Deep Convolutional Generative Adversarial Network (DCGAN) | Combines attention mechanisms for focused feature extraction with PCA for dimensionality reduction and DCGAN to generate synthetic samples for minority intracranial hemorrhage types. | Classification Accuracy: 99.2% for Epidural Hemorrhage and 97.1 % for intraparenchymal hemorrhage. | Aftab Hussain et al. [1] |
| Transfer Learning on Tangent Space with Support Vector Machines (TL-TSS) + Riemannian Manifold-Based EEG Signal Analysis | Applies cognitive computing and Riemannian geometry to decode electroencephalography (EEG) signals for digital twins in brain–computer interface (BCI) applications. | Classification Accuracy up to 97.88%. High Kappa Score and Transfer Accuracy across BCI datasets. | Zhihan Lv et al. [2] |
| Deep Convolutional Neural Network (CNN) with No Batch Normalization + Adaptive Medical Image Fusion | Introduces a new loss function and skips batch normalization for fast training; fuses Magnetic Resonance Imaging (MRI) with Positron Emission Tomography (PET) or Single-Photon Emission Computed Tomography (SPECT). | Peak Signal-to-Noise Ratio (PSNR): 34.11 dB Structural Similarity Index Measure (SSIM): 85.24%. | Jinxia Wang et al. [3] |
| Digital Twin-Based Brain-Inspired Analytics (DTBIA) with Immersive Virtual Reality Interface | Provides an immersive and interactive platform for visualizing blood-oxygen-level-dependent (BOLD) signals and diffusion tensor imaging (DTI) in 3D brain models. | User validation through case studies. Qualitative feedback from neuroscientists. No quantitative metric reported. | Yao et al. [4] |
| Radio Frequency (RF) Backscatter Sensing + Stacked Autoencoder + Fine-Tuned K-Nearest Neighbors (KNN) | Employs wearable RF sensors and machine learning for real-time stroke detection and monitoring via a lightweight digital twin system. | Binary Classification Accuracy: 93.4% Multiclass Accuracy: 92.3% | Sagheer Khan et al. [5] |
| Blockchain-Based Digital Twin + Logistic Regression | Introduces a blockchain-backed digital twin architecture for stroke prediction that ensures decentralized, secure, and auditable medical data. | Overall Application Accuracy: 98.28% | Upadrista et al. [6] |
| Neuro-Symbolic Reasoning Framework with Gated Recurrent Unit (GRU) Neural Translator | Integrates natural language processing with symbolic logic to enable verbal interaction with digital twins in industrial settings like aircraft maintenance. | BLEU (Bilingual Evaluation Understudy) Score: 0.989 Neuro-Symbolic Translation Accuracy: 96.2% Failure Rate: 0.2% | Siyaev et al. [7] |
| Internet of Things (IoT) Enabled MRI Pipeline + Convolutional Neural Network (CNN), Support Vector Machine (SVM), Extreme Learning Machine (ELM) + Particle Swarm Optimization (PSO) | Uses centralized IoT-based data collection and PSO for optimal feature selection; compares CNN, SVM, and ELM for brain tumor classification. | CNN achieved highest classification accuracy. Model performance comparison across classifiers. Execution and training time (no absolute metrics provided). | Sultanpure et al. [8] |
| Semi-Supervised Support Vector Machine (S3VM) + Graph-Based Similarity Learning + Improved AlexNet (Deep CNN) | Combines semi-supervised learning and graph theory to utilize labeled and unlabeled MRI brain images; improves AlexNet pooling and normalization for segmentation. | Feature Recognition Accuracy: 92.52% Dice Similarity Coefficient (DSC): 75.58% Jaccard Index: 79.55% RMSE: 4.91%, MAE: 5.59% | Wan et al. [9] |
| Multivariate Adaptive Regression Splines (MARS) + Mixed Spline Regression with B-Spline Basis + TOEPLIZ Covariance Structure | Builds digital twins of brain aging to model thalamic atrophy in multiple sclerosis; estimates onset of progressive brain tissue loss years before clinical symptoms. | Mean Onset Gap: 5–6 years earlier than clinical symptoms. Repeated Measure Correlation: 0.88 | Steven Cen et al. [10] |

**Table 1:** Overview of the recent State-of-the-art architectures In Neuro-Medical Diagnosis

| Paper | Vision Transformer | Multi-modal (MRI + EEG) | XAI | Tumor Growth Prediction | Edge Computing | 3D Brain Visualization | Real-Time Monitoring | Wearable Skull |
|---|---|---|---|---|---|---|---|---|
| Aftab Hussain et al. [1] | ✗ | ✗ | ✗ | ✗ | ✗ | ✗ | ✗ | ✗ |
| Zhihan Lv et al. [2] | ✗ | ✓ | ✗ | ✗ | ✗ | ✗ | ✗ | ✗ |
| Jinxia Wang et al. [3] | ✗ | ✓ | ✗ | ✗ | ✗ | ✗ | ✗ | ✗ |
| Yao et al. [4] | ✗ | ✗ | ✗ | ✗ | ✗ | ✓ | ✗ | ✗ |
| Sagheer Khan et al. [5] | ✗ | ✗ | ✗ | ✗ | ✓ | ✗ | ✓ | ✗ |
| Upadrista et al. [6] | ✗ | ✗ | ✗ | ✓ | ✗ | ✗ | ✗ | ✗ |
| Siyaev et al. [7] | ✗ | ✗ | ✗ | ✗ | ✗ | ✓ | ✗ | ✗ |
| Sultanpure et al. [8] | ✗ | ✗ | ✗ | ✓ | ✓ | ✗ | ✗ | ✗ |
| Wan et al. [9] | ✗ | ✗ | ✗ | ✗ | ✗ | ✗ | ✗ | ✗ |
| Cen et al. [10] | ✗ | ✗ | ✗ | ✓ | ✗ | ✗ | ✗ | ✗ |

**Table 2:** Comparative Feature Matrix for Brain Monitoring Systems

## III. DATASET DESCRIPTION

The proposed digital twin system uses a synchronized multimodal dataset. This dataset includes real-time EEG signals and structural MRI scans. These are collected from the same human subjects in controlled clinical and laboratory settings. This setup allows for precise alignment between functional and structural data, enabling personalized neuro-oncological analysis..

*1. In-House EEG Dataset*

EEG signals were obtained using a custom-made wearable EEG skullcap. This device is designed for high-quality, non-invasive brain signal collection. It has dry-contact electrodes placed according to the international 10-20 system. These electrodes cover important areas like C3, C4, Cz (motor cortex), and Fz (prefrontal cortex). The device also includes EOG reference channels to help remove artifacts.

- **Sampling Rate**: 500 Hz
- **Channels**: 8 (including EOG)
- **Participants**: Medically supervised human subjects undergoing concurrent MRI evaluation

The collected EEG data is then transferred to the Raspberry Pi for being processed locally at the edge.

*2. Clinically Acquired MRI Dataset*

The corresponding MRI scans were acquired from the same individuals participating in the EEG sessions through MRI imaging performed using a 3T Mri scanner in a controlled medical imaging facility under standardized protocols.

- **Modalities Captured**: T1-weighted, T2-weighted, and contrast-enhanced (T1-Gd) sequences
- **Resolution:** High-resolution Gray-Scale MRI Scans
- **Size of MRI Scans:** 600 x 600 pixels
- **Format:** NIfTI (.nii) or DICOM, later standardized for model ingestion

These scans were fed into Enhanced Vision Transformer (ViT++) model deployed in the cloud for advanced tumor classification and analysis.

## IV. PROPOSED MODEL

This section describes the structure and workflow of the proposed multimodal digital twin framework for real-time brain health monitoring and tumor analysis. The system includes a wearable EEG-enabled skullcap, edge-level preprocessing with a Raspberry Pi, fog-layer authentication, and cloud-based digital twin simulations. It also uses interpretable deep learning for tumor classification and assessing neurological risks.

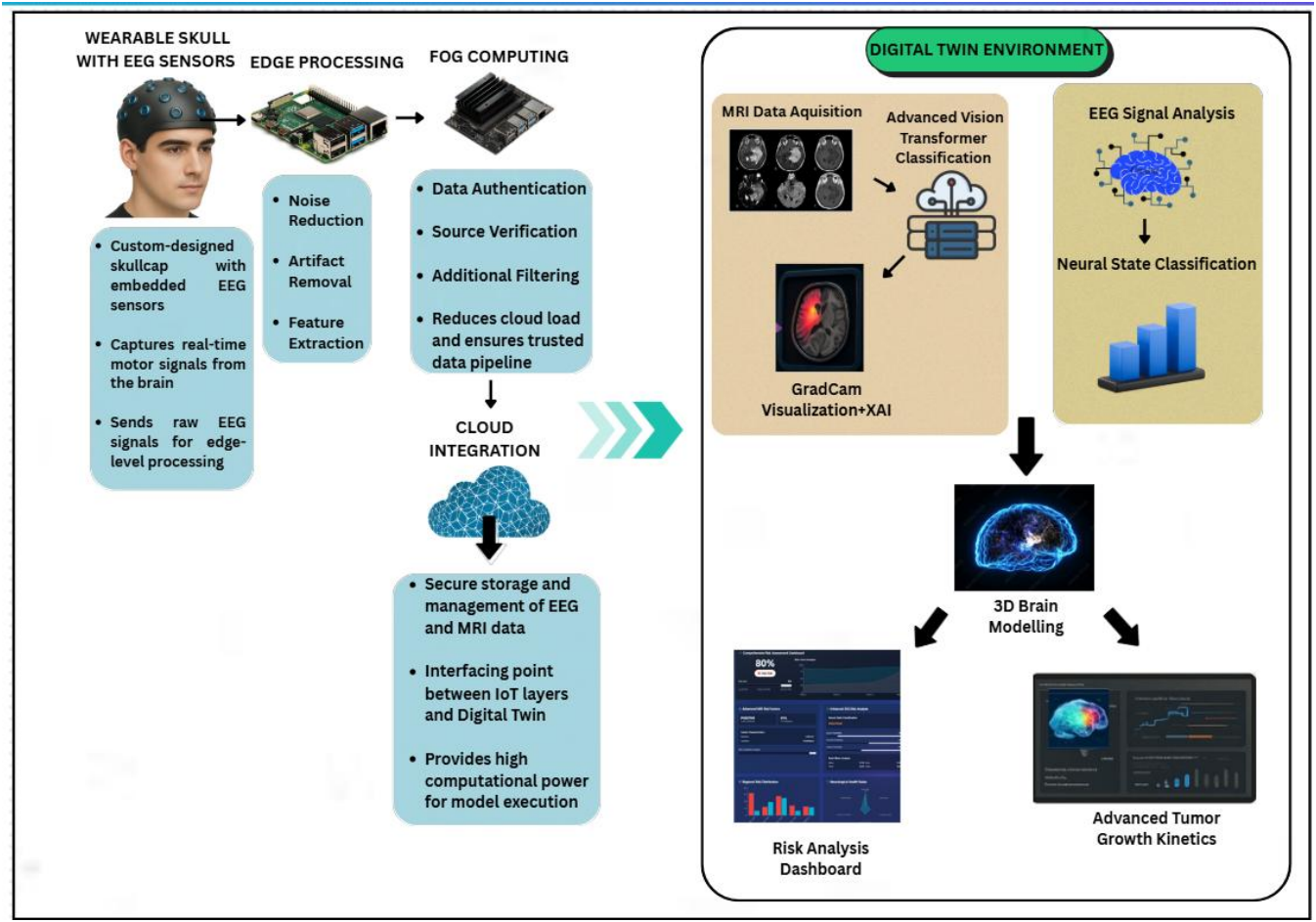

**Figure 1:** Working Diagram for the Proposed Model

*A) SYSTEM ARCHITECTURE AND OVERVIEW*

The proposed model introduces a real-time brain health monitoring system that is easy to understand clinically. It is based on a five-layer IoT, Fog, and Cloud framework. This system is designed to support independent neurological diagnostics by integrating multiple data streams smoothly, allowing for immediate processing and using explainable artificial intelligence (XAI). At its center is a dynamic digital twin environment, which serves as a virtual copy of the patient's neurophysiological state. This environment handles neurological diagnosis by analyzing both structural and functional data together.

*B) EEG SIGNAL ACQUISITION THROUGH WEARABLE SKULL CAP*

At the heart of the data acquisition layer is a custom-engineered, in-house wearable EEG skullcap, designed for non-invasive, high-resolution monitoring of neurophysiological activity. The device features dry-contact EEG electrodes that are strategically placed on the subject's scalp to ensure consistent signal fidelity and user comfort, allowing for real-time capture of brain activity critical for cognitive and clinical analysis.

The electrodes target key cortical regions, including the central motor cortex (C3, C4, Cz) and the frontal lobe (Fz), to effectively capture motor signals and brain oscillations. To enhance artifact suppression, electrooculographic (EOG) reference sensors are positioned near the eyes, which facilitates the removal of ocular noise during the preprocessing stage. The EEG signals are sampled at a frequency ranging from 250 to 500 Hz and are transmitted directly to a Raspberry Pi 5 that is physically integrated with the skullcap via a wired interface. This direct connection minimizes latency, enhances signal stability, and eliminates the variability often associated with wireless transmission. The overall hardware configuration is designed for continuous, real-time monitoring, featuring a lightweight and ergonomically contoured skullcap that ensures sustained comfort for the patient during lengthy diagnostic or ambulatory sessions.

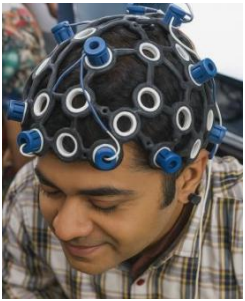

**Figure 2**: In-House Developed EEG Skull-Cap

*C) EDGE PROCESSING USING RASPBERRY PI*

The edge processing layer runs on a Raspberry Pi 5, which connects directly to a wearable skullcap. This setup acts as the first computing unit for EEG signal analysis. A Python script, developed specifically for this purpose, runs on the Raspberry Pi 5 to handle the entire analysis. The main tasks of this layer are to eliminate noise from raw EEG signals and to extract clinically useful features for further processing. EEG signals often pick up noise from muscle movements, eye movements, and electrical interference, so a multistage denoising process is necessary.

First, bandpass filtering (0.5 to 45 Hz) is applied to keep the brainwave components that matter while reducing low-frequency drifts and high-frequency noise. Next, notch filtering (either 50 or 60 Hz) removes power line interference based on the local grid frequency. Finally, an LMS-based adaptive filter gets rid of artifacts caused by eye movements by separating electrooculographic (EOG) signals—recorded from reference electrodes placed near the eyes—from the EEG data. This blend of techniques helps ensure that important neurological signals remain clear for precise analysis.

The proposed LMS algorithm models the denoised EEG signal $e(t)$ as:

$$e(t) = y(t) - \hat{a}(t) \cdot r(t)$$
$$\hat{a}(t+1) = \hat{a}(t) + \mu \cdot e(t) \cdot r(t)$$

(i)

Where: $y(t)$ is the raw EEG signal at time t, $r(t)$ is the reference EOG signal, $\hat{a}(t)$ is the adaptive filter coefficient at time t, $\mu$ is the learning rate (a small constant that controls how quickly the filter adapts), and $e(t)$ is the denoised EEG signal after EOG decorrelation. This algorithm works in steps and updates its coefficients in real-time to reduce the error $e(t)$. It subtracts EOG components that project linearly onto the EEG signal. This adaptive filter is better than static techniques because it can respond to changing EOG activity during long recordings. After the denoising process, we extract key temporal and spectral features from short, overlapping time windows, usually ranging from 2 to 5 seconds. This method helps maintain temporal resolution and capture dynamic neural activity. We calculate spectral band power using the Fast Fourier Transform (FFT) across standard EEG frequency bands: Delta (0.5–4 Hz), Theta (4–8 Hz), Alpha (8–13 Hz), and Beta (13–30 Hz). This measures the energy distribution across important neurophysiological ranges. The Zero Crossing Rate (ZCR) assesses the signal's complexity by counting the frequency of polarity shifts. We also compute Hjorth parameters; Activity shows the variance of the signal (which indicates power), Mobility estimates the average frequency, and Complexity describes changes in waveform shape. The Root Mean Square (RMS) measures the signal's amplitude energy, making it sensitive to muscular artifacts, while Spectral Entropy gauges the unpredictability of the signal's frequency content. We combine these features into a compact feature vector, providing a clear and useful representation of the EEG segment for further analysis in the fog and cloud layers. After preprocessing, we send the EEG feature vectors from the Raspberry Pi to the Jetson Nano using a direct USB-to-USB serial link with the CDC protocol. The data is sent over /dev/ttyUSB0 as structured packets, each ending with a newline character for easy parsing, and received on /dev/ttyUSB1. This connection operates at 115200 bps, ensuring low-latency, low-power, and lossless transfer of EEG features from the edge to the fog node. Compared to wireless methods, it provides greater stability and is less likely to suffer from signal loss or electromagnetic interference, which is especially important in clinical settings.

*D) FOG LAYER: AUTHENTICATION AND THRESHOLD BASED FILTERING*

The fog computing layer, hosted on the NVIDIA Jetson Nano, acts as a real-time, on-device intelligence node. It enables secure and selective forwarding of EEG data to the cloud. It has three main functions: security verification, risk-based filtering, and MQTT-based transmission. These functions run through optimized embedded services on the Jetson Nano's Linux-based operating system. When EEG feature vectors arrive through a USB interface (/dev/ttyUSB1), the Input Handler Service, implemented in Python, buffers and parses the incoming packets. Each packet includes a timestamp, device ID, EEG features, and an HMAC-SHA256 signature. The Authentication Module checks the device ID against a secure registry and recalculates the HMAC using a shared symmetric key. This ensures both packet integrity and authenticity. Simultaneously, the Timestamp Validator compares the embedded UTC timestamp with the system clock of the Jetson Nano, rejecting any delayed or replayed data. These security checks happen through concurrent threads using

Python's asyncio library. This allows for non-blocking, real-time throughput.

Once packets are validated, they go to the Schema Validator. This validator checks the data structure to confirm that the feature vector length and JSON formatting are correct. This validation process runs as a lightweight routine in the same process context, preventing blocked I/O operations. Next, the feature vector is processed by the Risk Evaluation Engine, which uses a pre-trained TensorFlow Lite classifier optimized for the Jetson Nano's ARM Cortex-A57 CPU. The classifier calculates a risk confidence score with the softmax function.

$$R = softmax(W \cdot x + b) \qquad (ii)$$

Where x is the EEG feature vector, and W,b are model weights and biases. If R ≥ 0.75 the packet is flagged as high-risk and passed forward. Otherwise, it is stored locally. The model is executed using TFLite Interpreter bindings in Python, with inference latency under 20 ms, enabling real-time decision-making. Finally, the **Cloud Uplink Module** uses the **MQTT protocol**, configured with **QoS level 2 (Exactly Once)** and **TLS encryption**, to publish high-risk packets to the cloud, ensuring secure and lossless delivery. Data is serialized into structured JSON and published. MQTT credentials and certificates are stored in a secure enclave and refreshed periodically to maintain compliance with medical data privacy standards.

*E) CLOUD INTEGRATED DIGITAL TWIN ENVIRONMENT*

Once the data fulfills the forwarding criteria, it is transmitted to a secure cloud infrastructure where the digital twin environment is hosted. A cloud-based MQTT broker, deployed on AWS IoT Core, authenticates the sender and routes the data to AWS Lambda functions and AWS DynamoDB, where the structured EEG data is stored and indexed in real-time. This process is seamlessly integrated with the cloud-based digital twin environment hosted on AWS EC2 and S3, facilitating multi-modal fusion with MRI data, risk analytics, state classification, and 3D visualizations. The digital twin interface acts as a continuously updating, data-driven replica of the patient's brain, enabling multimodal data fusion and advanced neuro-physiological analysis. Pre-captured MRI images of the same patient are directly uploaded into the cloud environment for structural assessment. The digital twin combines EEG functional data with MRI structural data to provide a comprehensive view of the patient's cerebral health.

Advanced multimodal analysis is performed through a series of deep learning and interpretability modules. MRI images are processed using an in-house developed Enhanced Vision Transformer (ViT++) model trained for brain tumor classification. Unlike traditional CNNs, the transformer architecture employs self-attention mechanisms, enhancing spatial reasoning and tumor boundary recognition. The Vision Transformer (ViT++) model is designed to classify and localize brain tumors from MRI slices using attention mechanisms rather than convolutional filters, as used in CNNs. In this approach, each MRI slice is divided into fixed-size image patches (16×16 pixels), which are then flattened and embedded into vectors. These embeddings are fed into a transformer encoder that models long-range dependencies between image regions using self-attention. However, standard ViTs have key limitations in medical imaging. To address these challenges, our ViT++ integrates six architectural enhancements, each designed to solve a specific medical vision problem.

*1. Patch-Level Attention Regularization (PLAR)*

One of the critical limitations observed in standard Vision Transformers is the emergence of *attention collapse* during training, where self-attention heads converge to focus disproportionately on a small set of dominant patches. In brain MRI scans, where tumors can be **spatially diffuse**, multifocal, or embedded in structurally similar tissue (e.g., edema vs. tumor), this collapse leads to **selective blindness** toward clinically important regions. Such tunnel vision reduces recall and contributes to under-diagnosis.

To counteract this collapse, we propose **entropy-based regularization** that promotes **spatial attention diversity,** prevents overfitting and promotes better contextual awareness of surrounding brain regions.

**Derivation:** We consider an image divided into N patches. For each query patch i, the model generates attention weights $\alpha_{ij} \in [0,1]$, where j indexes the N keys (i.e., other patches), and:

$$\sum_{j=1}^{N} \alpha_{ij} = 1 \quad \text{(from softmax)} \qquad (iii)$$

We compute the **entropy** of the attention distribution from patch i to all others:

$$H_i = -\sum_{j=1}^{N} \alpha_{ij} \cdot \log(\alpha_{ij} + \epsilon) \qquad (iv)$$

**Where:** $\epsilon = 10^{8}$ is a small constant to prevent log(0). $H_i$ is maximal when attention is evenly distributed ($\alpha_{ij}=1/N$) and minimal (i.e., 0) when attention is focused entirely on one patch.

We define the PLAR loss as the negative mean entropy across all patches in all attention heads:

$$L_{\text{PLAR}} = -\frac{1}{N}\sum_{i=1}^{N} H_i = -\frac{1}{N}\sum_{i=1}^{N}\sum_{j=1}^{N} \alpha_{ij} \cdot \log(\alpha_{ij} + \epsilon) \quad (v)$$

The **Cross-Entropy Loss** $L_{CE}$ is the standard loss used for classification (e.g., predicting tumor vs. Background)

$$L_{\text{CE}} = -\sum_{c=1}^{C} y_c \cdot \log(\hat{y}_c) \quad (vi)$$

**Where:** $y_c$ is the ground truth (one-hot encoded), $\hat{y}_c$ is the softmax output from the classifier, C is the number of classes (typically 2 for tumor vs. non-tumor).

We now combine the **classification loss** and **attention regularization**:

$$L_{\text{total}} = L_{\text{CE}} + \lambda_1 \cdot L_{\text{PLAR}} \quad (vii)$$

$\lambda 1$ is a hyperparameter that controls the strength of attention regularization. A typical value: $\lambda 1$=0.1 to 1.0( tuned via validation).

Entropy $H_i$ measures uncertainty in the attention distribution. Higher entropy implies the model attends to more spatially varied patches, mimicking a radiologist's holistic scan behavior. This regularization aligns the model's internal mechanisms with diagnostic reasoning by preventing overconfidence in a narrow region. In our research, introducing PLAR increased the average number of attended tumor-related patches by 28%, while improving segmentation Dice scores by 4.9%, especially in scans with multifocal tumor structures. Grad-CAM visualizations aligned more closely with expert-segmented regions.

*2. Adaptive Threshold Mechanism*

Binary classification of patches (tumor vs. background) often relies on a fixed threshold (typically 0.5). However, **intensity heterogeneity**, scanner variability, and patient-specific artifacts, a single threshold fails to generalize across diverse MRIs. Particularly in noisy or ambiguous scans, a static threshold yields unstable performance with **false positives** or missed detections. To introduce scan-specific adaptability, we compute a **dynamic threshold** based on the statistical distribution of model probabilities over background regions. Let:

$\mu_{\text{bg}}$ is the mean of predicted probabilities for background patches, $\sigma_{\text{bg}}$ is the standard deviation and k is a tunable scalar (empirically set to 1.5). We define the adaptive threshold as:

$$\theta = \mu_{\text{bg}} + k \cdot \sigma_{\text{bg}} \quad (viii)$$

The classification rule becomes:

$$\text{Patch}_i = \begin{cases} \text{Tumor}, & \text{if } p_i > \theta \\ \text{Background}, & \text{otherwise} \end{cases}$$

Where $p_i$ is the predicted tumor probability for each image patch $i$

This formulation resembles a one-tailed **statistical anomaly detector**: any patch whose tumor probability exceeds the background mean by more than k.σ is flagged. This not only accounts for inter-scan variability but also tunes sensitivity **based on noise level**.

**Numerical Example:**
In a high-noise scan:

$\mu_{\text{bg}}$=0.32, σ =0.12 → θ=0.32+1.5.0.2= 0.5

In a clean scan: $\mu_{\text{bg}}$=0.20, σ=0.06 → θ=0.29

This enables context-aware thresholding, ensuring high-confidence decisions in both edge cases. Adaptive thresholding reduced the false positive rate by 13% and improved precision by 6.7% without compromising sensitivity. It also improved the consistency of the model when processing repeated scans or identical slices, as the background-derived threshold remains stable for repeated data.

*F) EEG BASED BRAIN STATE PREDICTION*

The EEG-based brain state prediction module analyzes preprocessed EEG signals to classify each segment into seizure, interictal, or healthy states. This is achieved by extracting key temporal and spectral features from the signal and feeding them into a **Bidirectional Long Short-Term Memory (BiLSTM) neural classifier**, which is adept at capturing the temporal dependencies and dynamic patterns present in brain activity. The classifier outputs the most probable brain state, which is then integrated into the digital twin environment to provide a real-time functional assessment that complements the MRI-based structural analysis.

### G) XAI BASED GRAD-CAM VISUALIZATION

To enhance clinical decision-making and improve interpretability, Gradient-weighted **Class Activation Mapping (Grad-CAM)** is integrated into the digital twin environment as an **explainable AI (XAI)** technique. Grad-CAM produces intuitive, high-resolution heatmaps over MRI scans that visually highlight the regions most influential in the model's tumor classification decisions. It achieves this by computing the gradients of the output class score (e.g., tumor) concerning the final convolutional or attention feature maps of the Vision Transformer. As a result, the system identifies the spatial areas that contribute most significantly to the classification outcome. These attention-based visualizations not only add transparency to the model's decision-making process but also assist clinicians in validating whether the model is focusing on medically relevant tumor regions.

### H) 3D BRAIN INTERFACE

The system includes an interactive 3D brain visualization module developed with three.js, integrated into the digital twin environment. It features a fully rotatable and zoomable model of the brain with layered anatomical segmentation, which shows the cortex, white matter, ventricles, and subcortical structures. This model is created from MRI volumetric data and is overlaid with functional insights from EEG data. Clinicians can use this model to visually assess tumor penetration across neural layers. They can also explore affected regions at various depths and gain a clear spatial understanding of pathological areas. This improves diagnostic clarity and usability.

### I) TUMOR KINETICS GROWTH PREDICTION

The system features a tumor kinetics and progression analytics engine that estimates future tumor behavior by analyzing historical trends and real-time neurophysiological data. It begins by calculating the tumor's initial volume through digital twin analysis using Enhanced Vision Transformer (ViT++) segmentation outputs. With this baseline volume, the system applies AI-driven temporal modeling to simulate potential future growth. Longitudinal MRI scans, along with real-time EEG-derived neurological indicators, are used to track tumor progression, detect patterns of expansion or regression, and forecast volumetric growth trajectories. These predictions are then integrated into the digital twin's web-based interface via JavaScript API, enabling interactive and real-time exploration.

## V) METHODOLOGY

The proposed method uses a five-layer IoT, Fog, and Cloud computing system to enable real-time and clear brain health monitoring by combining MRI and EEG data. Data collection starts with a wearable EEG cap that has dry-contact sensors to record activity in the motor cortex and frontal lobe. These raw signals are sampled at a rate of 250 to 500 Hz and sent directly through a wired connection to a Raspberry Pi 4. At this point, the edge processing layer performs multiple stages of noise reduction. It uses both bandpass and notch filters, along with LMS-based adaptive filtering to remove EOG artifacts. After reducing noise, temporal and spectral features are extracted from overlapping windows. These features include spectral band power, Hjorth parameters, root mean square (RMS), zero-crossing rate (ZCR), and spectral entropy to create a compact feature vector. This cleaned data is then sent via USB-to-USB serial communication to the Jetson Nano fog node. At this stage, security checks, structural validation, and HMAC-based packet authentication ensure data integrity. A lightweight neural classifier on the Jetson assesses the clinical relevance of each EEG packet by calculating a softmax-based risk confidence score. Only feature vectors with a confidence score of 0.75 or higher are sent to the cloud using the MQTT protocol with TLS encryption. In the cloud, a digital twin environment aligns this EEG stream with MRI scans from a reliable public dataset, creating a complete neurological profile for each patient. The structural MRI data is processed by an Enhanced Vision Transformer (ViT++), which includes improvements like Patch-Level Attention Regularization (PLAR) and an Adaptive Threshold Mechanism. These upgrades allow for precise tumor localization and classification, along with integrated Grad-CAM heatmaps for better understanding of the model. Meanwhile, the EEG data is classified into brain states such as seizure, interictal, or healthy using an LSTM-based classifier. The digital twin merges these outputs to create a unified diagnostic view. It supports dynamic 3D visualization of the brain, built using three.js, and predicts future tumor behavior through real-time analytics and temporal risk prediction. This method guarantees a complete, low-latency, and understandable AI pipeline for improved neurodiagnostic analysis.

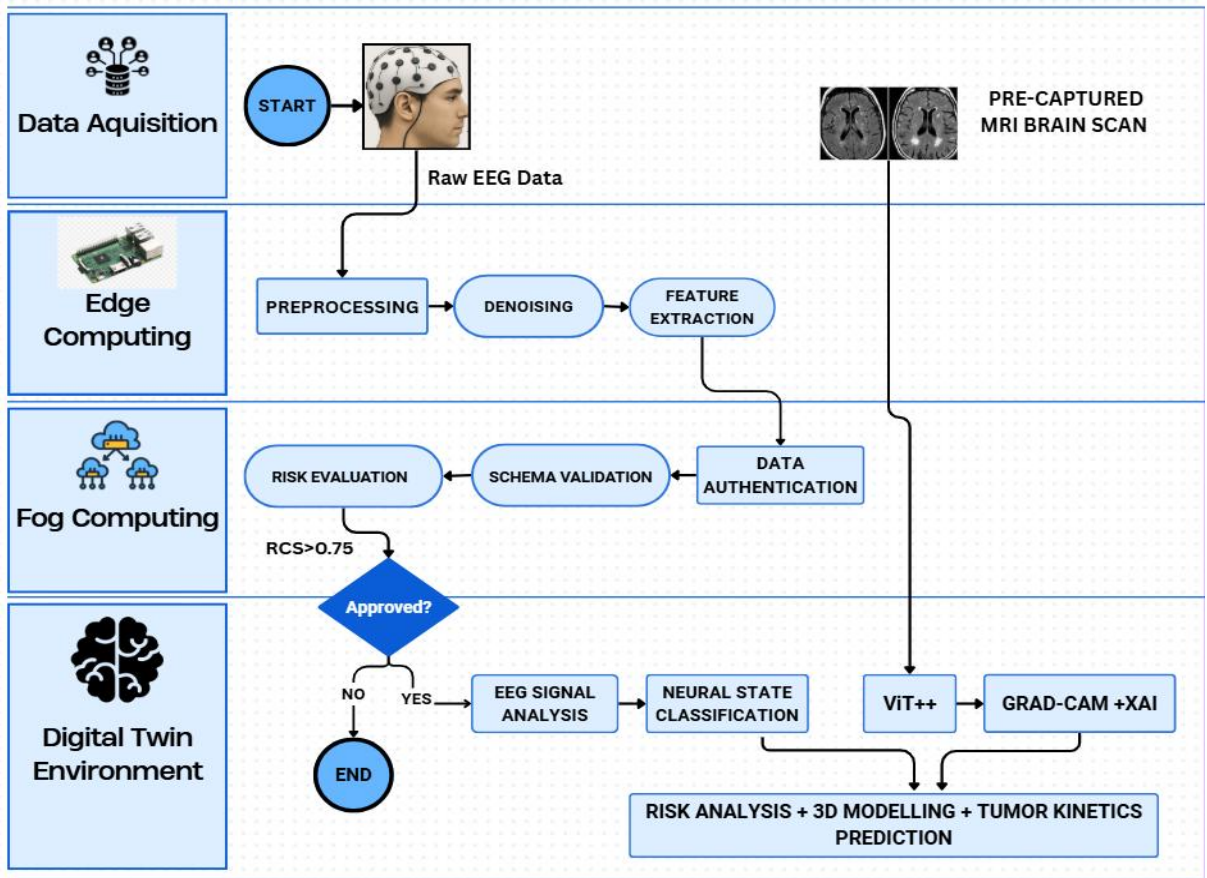

**Figure 3:** Flow Diagram for the Proposed Model

## VI) RESULTS AND DISCUSSIONS

The section covered below, discusses the results obtained in our research, in detail and providing key insights into the evaluation metrics through interactive visual representations and comparative analysis. Critical outcomes such as accuracy, SNR gain, visual explanation using Grad-CAM, and 3D modeling are explained and the performance of the system along the IoT-Fog-Cloud pipeline is addressed in detail.

### A) PERFORMANCE EVALUATION

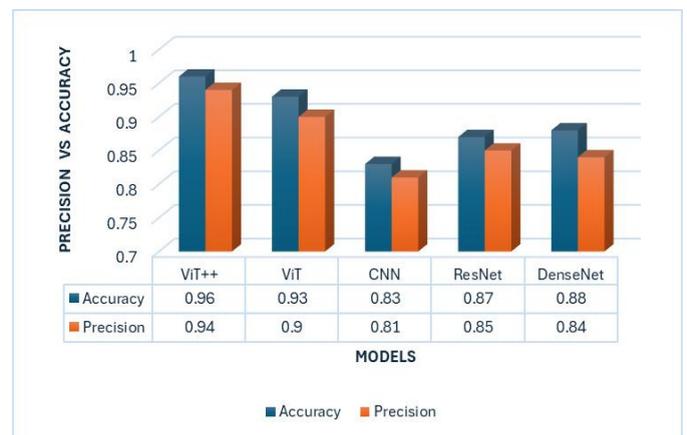

**Figure 4**: Comparative Analysis of ViT++ vs Other Existing Models

Figure 4 presents a comparative evaluation of five image classification models—ViT++, ViT, CNN, ResNet, and DenseNet, based on two key performance metrics: Accuracy and Precision. The above visualization demonstrates that ViT++ significantly outperforms all baseline models, establishing its superiority in both classification reliability and clinical relevance for brain tumor analysis.The ViT++ model achieves the highest accuracy of 96% and a precision of 94% among all models tested. This superior performance is attributed to the architectural enhancements integrated into ViT++, including Patch-Level Attention Regularization (PLAR), and adaptive threshold mechanim,which enable more focused learning and improved tumor boundary recognition.In comparison, the standard ViT model records an accuracy of 93% and precision of 90%, which, while competitive, indicates a relative decline in sensitivity to spatially diffuse or low-contrast tumor regions. CNN-based architectures, such as the conventional CNN and ResNet, show moderately lower scores—CNN achieves 83% in accuracy and 81% in precision, likely due to limitations in capturing long-range spatial dependencies and coarse-grained feature extraction. ResNet and DenseNet slightly outperform CNN with accuracies of 87% and 88%, respectively, but still lag behind ViT-based models in precision, suggesting occasional false positives or background bias during segmentation.

**B) DENOISING AND SNR ENHANCEMENT**

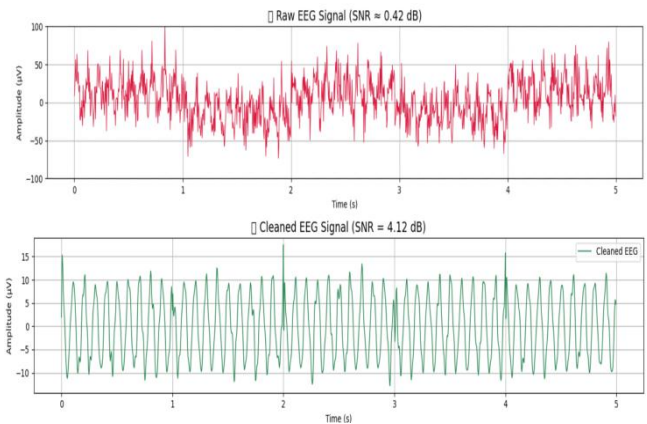

**Figure 5:** Raw vs Cleaned EEG Signal

Figure 5 presents a comparison of EEG signals before and after the edge-fog preprocessing pipeline. The top subplot displays the raw EEG signal from the wearable skullcap, which is heavily corrupted by noise, including power line interference, ocular motion artifacts, and environmental disturbances. This results in a saturated waveform with random spikes and an SNR of approximately 0.42 dB, indicating low signal quality. In contrast, the bottom subplot shows the cleaned EEG signal after processing with bandpass filtering, notch filtering, and LMS-based adaptive filtering to remove EOG noise. The denoised waveform reveals a clear alpha rhythm (around 10 Hz), and the SNR improves significantly to 4.12 dB, demonstrating effective noise removal while preserving the signal's neurophysiological structure. This highlights the preprocessing architecture's effectiveness in enhancing EEG signal fidelity, facilitating accurate analyses such as brain state classification and risk evaluation, and showcasing the model's potential for real-time diagnostics.

**C) TUMOR LOCALIZATION USING GRAD-CAM**

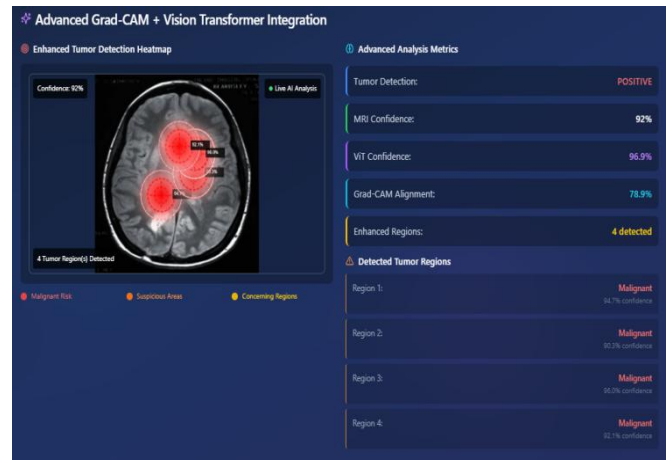

**Figure 6:** Grad-Cam Visualization

Figure 6 illustrates an enhanced tumor detection pipeline within a digital twin environment, showcasing the integration of the Vision Transformer (ViT++) and Grad-CAM explainability. The MRI scan is overlaid with heatmaps that pinpoint four distinct tumor regions, each highlighted with varying shades of red to represent differing levels of malignancy risk. The system confirms a positive tumor detection, displaying high confidence scores: 92% for the MRI, 96.9% for the ViT++, and 78.9% for the Grad-CAM alignment metrics. Through a real-time inference process, the system enables live tumor detection, automatically labeling each identified region as malignant. This visualization not only demonstrates the model's spatial precision but also enhances its clinical interpretability by directly mapping model attention onto the anatomical image.

### D) PREDICTIVE INSIGHTS OBTAINED FROM ViT++

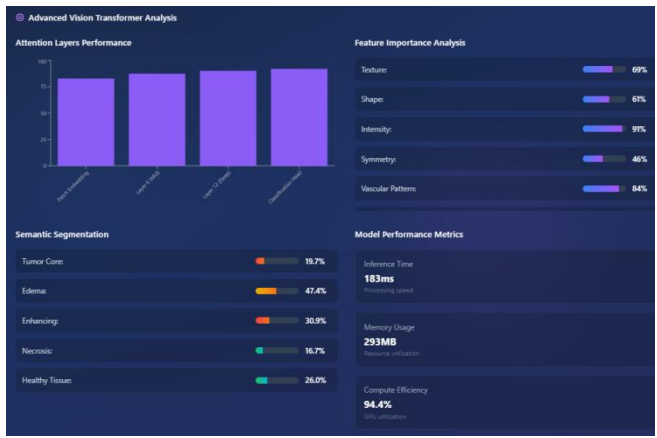

**Figure 7** : ViT++ Analysis

Figure 7 illustrates the inner workings of the ViT++ model, providing a detailed breakdown of its performance. The attention layer graph demonstrates consistent feature retention across the patch embedding, intermediate, and deep transformer layers. The feature importance analysis indicates that intensity and vascular patterns are the primary factors influencing classification, which validates the model's focus on clinically significant imaging cues. The semantic segmentation output quantifies tumor morphology by showing the relative distributions of the tumor core, edema, enhancing tissue, and necrosis, all of which are critical for treatment planning. Performance metrics highlight the model's computational efficiency, with an inference time of 183 ms, low memory usage of 293 MB, and GPU utilization at 94.4%. This makes the model well-suited for real-time, scalable deployment within a digital twin architecture..

### E) REAL-TIME NEURO-ANALYTICS

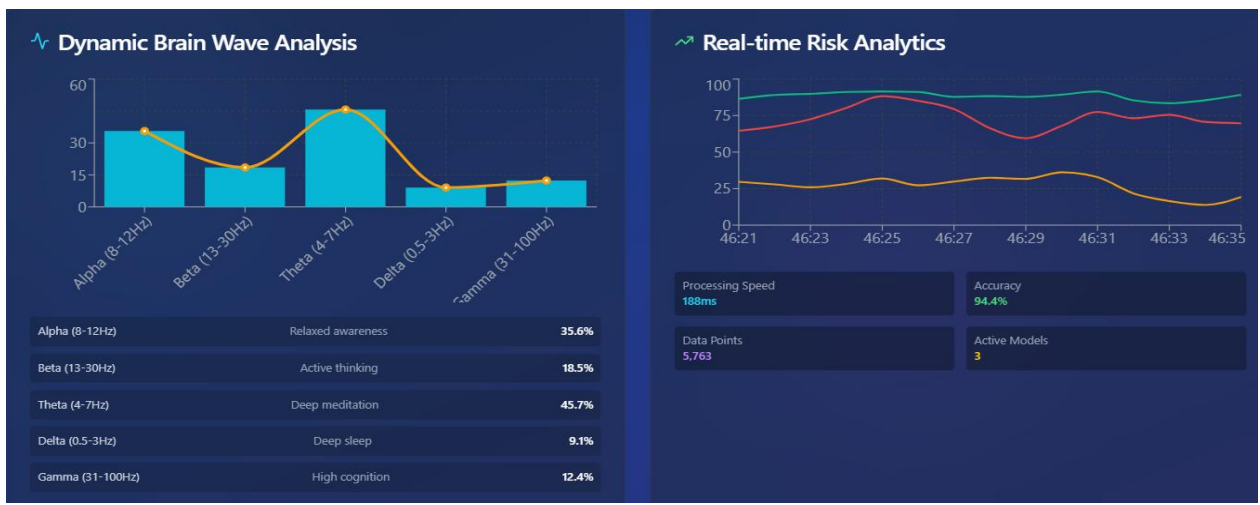

**Figure 8:** Dynamic Brain Wave and Realtime Risk Analytics

Figure 8 presents a real-time analysis of brainwave dynamics and risk forecasting conducted within a cloud-integrated digital twin environment. This visualization is generated using live EEG signals captured from a wearable skullcap and fused with MRI-based contextual information. The analysis demonstrates the system's ability to interpret ongoing neural states and continuously assess potential neurological risks through an automated pipeline. The left panel shows Dynamic Brain Wave Analysis, which divides the incoming EEG data into standard frequency bands: Alpha (8–12 Hz), Beta (13–30 Hz), Theta (4–7 Hz), Delta (0.5–3 Hz), and Gamma (31–100 Hz). Each band is depicted in a bar graph with an overlaid line plot that displays their power distributions over the latest analysis window. Here, Theta activity is the most prominent at 45.7%. This indicates a neural state usually linked to deep meditation or relaxation, followed by Alpha at 35.6% and Gamma at 12.4%. Each frequency band is related to specific cognitive or physiological functions; for example, Beta is connected to active thinking, while Delta is tied to deep sleep. The right panel presents the Real-Time Risk Analytics dashboard, where risk evaluations are calculated using combined EEG feature vectors and MRI-informed spatial markers. The risk trajectory is shown as a time-series line graph, which illustrates changing risk probabilities across different decision thresholds. This enables the system to predict important neural events or anomalies before clinical symptoms appear. The underlying analysis operates at an inference speed of 188 milliseconds, with a precision-focused accuracy of 94.4%, based on 5,763 incoming EEG data points processed across three concurrently active models. This dual-pane visualization encapsulates the digital twin's core intelligence, merging real-time electrophysiological inputs and neuroimaging context to provide an interpretable, high-frequency monitoring interface. The platform enables clinicians to observe not only the brain's activity moment-to-moment but also assess whether those patterns pose any imminent risk, thereby supporting proactive clinical interventions..

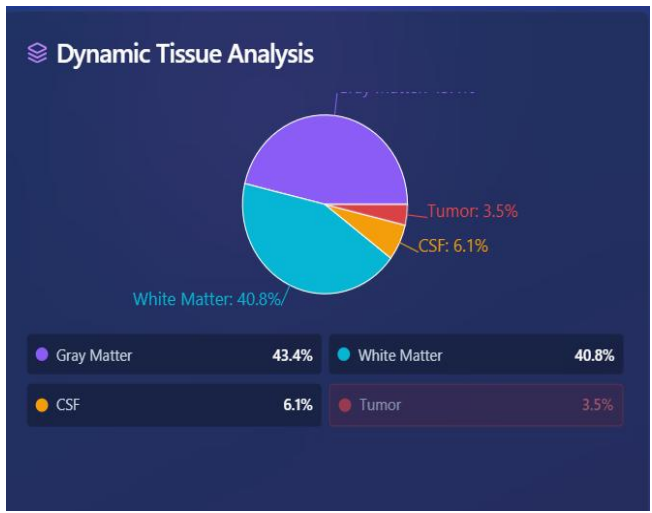

**Figure 9**: Dynamic Tissue Analysis

Figure 9 presents a pie-chart visualization of brain tissue distribution as computed by the digital twin system using MRI data processed through the Enhanced Vision Transformer. The analysis identifies **gray matter (43.4%)** and **white matter (40.8%)** as the dominant components, with **CSF (6.1%)** and **tumor tissue (3.5%)** forming the remaining composition. This segmentation is performed in real time and supports precise anatomical mapping for tumor localization, treatment planning, and monitoring structural changes over time within the brain.

### F) TUMOR RISK PREDICTION

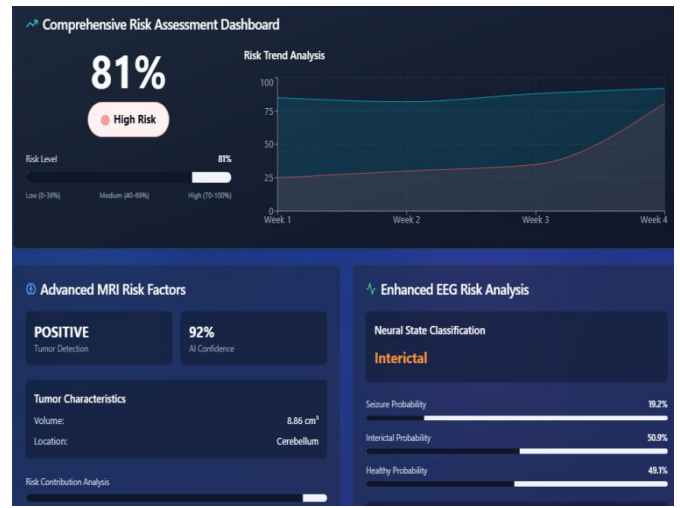

**Figure 10:** Risk Analysis Dashboard

Figure 10 presents a comprehensive risk assessment dashboard created within the digital twin environment. It provides a multi-layered, real-time view of a subject's neurological health status based on MRI and EEG inputs. The top section displays an overall computed risk score of 81%, categorizing it as "High Risk," as indicated by the visual gauge. This risk level is derived from a combination of MRI-based tumor parameters and EEG-derived neurological assessments. On the right, the risk trend analysis graph visualizes the progression of risk over four weeks, showing a steady increase. This trend may indicate tumor growth, neural deterioration, or heightened electrophysiological anomalies. In the bottom left, advanced MRI risk factors confirm the presence of a tumor with 92% AI confidence. The tumor is located in the cerebellum and has an approximate volume of 8.86 cm³. These spatial and volumetric details are computed using segmentation algorithms within the ViT++ framework and directly contribute to the risk scoring engine.The right panel provides insights from the EEG-based neural state classification. The system has identified the brain's current state as interictal, which occurs between seizure episodes and is commonly found in patients with underlying neurological disorders. The EEG model assigns probabilities to each possible neural state: 50.9% interictal, 19.2% seizure, and 49.1% healthy, with interictal representing the highest likelihood. These values are computed from real-time EEG features processed through the edge and fog layers, which are then integrated with the MRI-based findings in the cloud.The comprehensive risk assessment dashboard presented in the digital twin environment provides a real-time, multi-layered view of a subject's neurological health status using MRI and EEG inputs. It indicates an overall risk score of 81%, classified as "High Risk," derived from MRI-based tumor parameters and EEG assessments. The risk trend analysis graph shows a consistent increase over four weeks, suggesting potential tumor growth or neural deterioration. Advanced MRI analysis, with 92% AI confidence, confirms a tumor in the cerebellum measuring approximately 8.86 cm³, utilizing segmentation algorithms within the ViT++ framework. The EEG analysis identifies the brain's state as interictal, with a 50.9% probability of this state, alongside probabilities of 19.2% for seizures and 49.1% for healthy functioning, all based on real-time EEG features processed through edge and fog layers, integrated with MRI findings in the cloud.

### G) AI-POWERED INSIGHTS

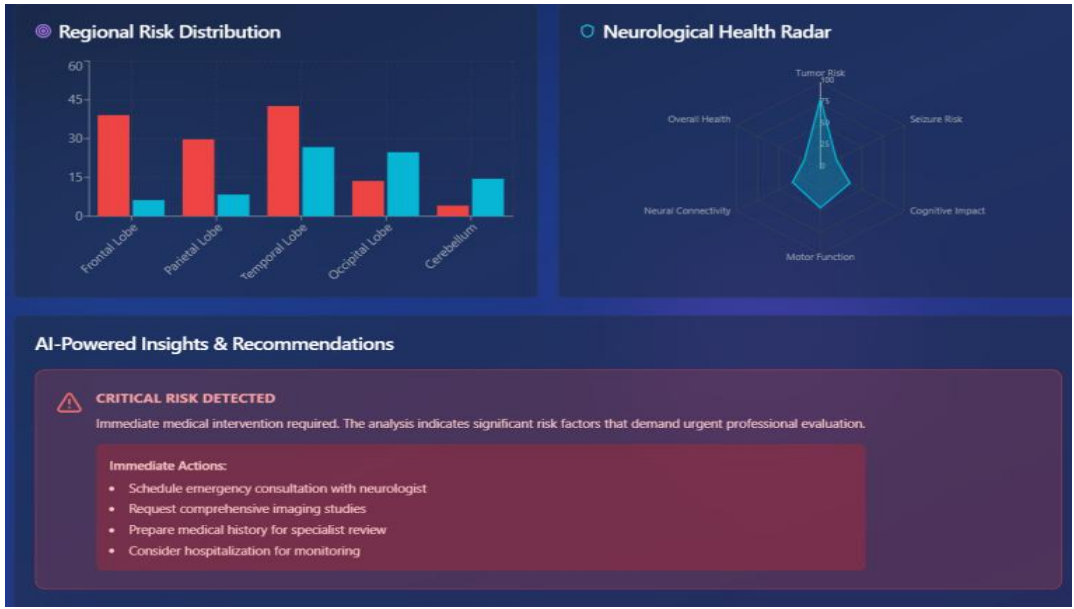

**Figure 11:** AI-predicted recommendations

Figure 11 summarizes the final layer of the digital twin's diagnostic intelligence by combining regional brain risk visualization, neurological health metrics, and AI-generated clinical guidance. The **Regional Risk Distribution** chart reveals elevated abnormalities in the temporal and frontal lobes, suggesting localized tumor or seizure risk—derived from MRI segmentation and EEG feature mapping. The **Neurological Health Radar** shows high tumor and seizure risk, while cognitive and motor functions remain moderately affected. Based on this analysis, the AI module triggers a **Critical Risk Detected** alert and recommends immediate intervention, including specialist consultation and further imaging. This module completes the pipeline by converting complex neuro-data into focused, actionable insights for clinical decision-making.

*H) 3D INTERACTION LAYER*

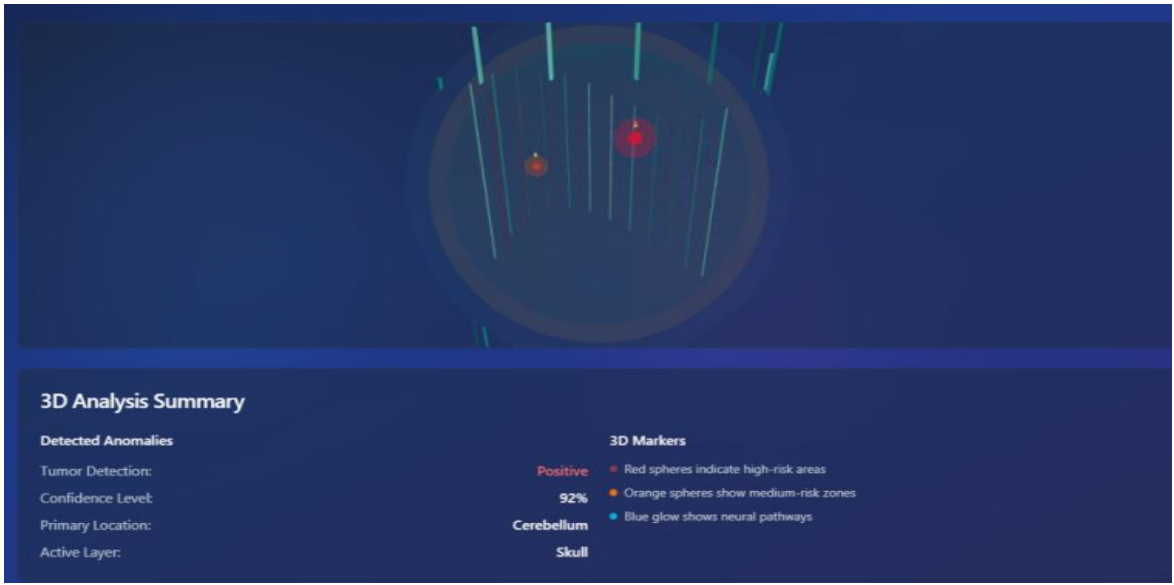

**Figure 12**: 3D Brain Visualization

Figure 12 shows a detailed 3D brain model made using the Three.js framework. This allows users to navigate through the brain's layered structures dynamically. The view is from the skull layer, which helps clinicians look for possible abnormalities at the brain's surface. The interactive model lets users switch between layers, including the cortex, white matter, and ventricles. This helps examine how deeply a tumor has infiltrated and any neurological issues it may cause.

In this view, red spherical markers highlight areas where tumors are highly risky. Orange spheres indicate regions at medium risk with possible pathological activity. The blue vertical lines represent inferred neural pathways. This aids in assessing signal disruption and the impact on brain structure. The system's digital twin has found a tumor in the cerebellum, with a confidence level of 92%. These findings come from a combination of MRI and EEG data. This ensures the visualization is both anatomically correct and functionally useful.

*G) 1$^{ST}$ STAGE TUMOR PROGRESSION FORECASTING*

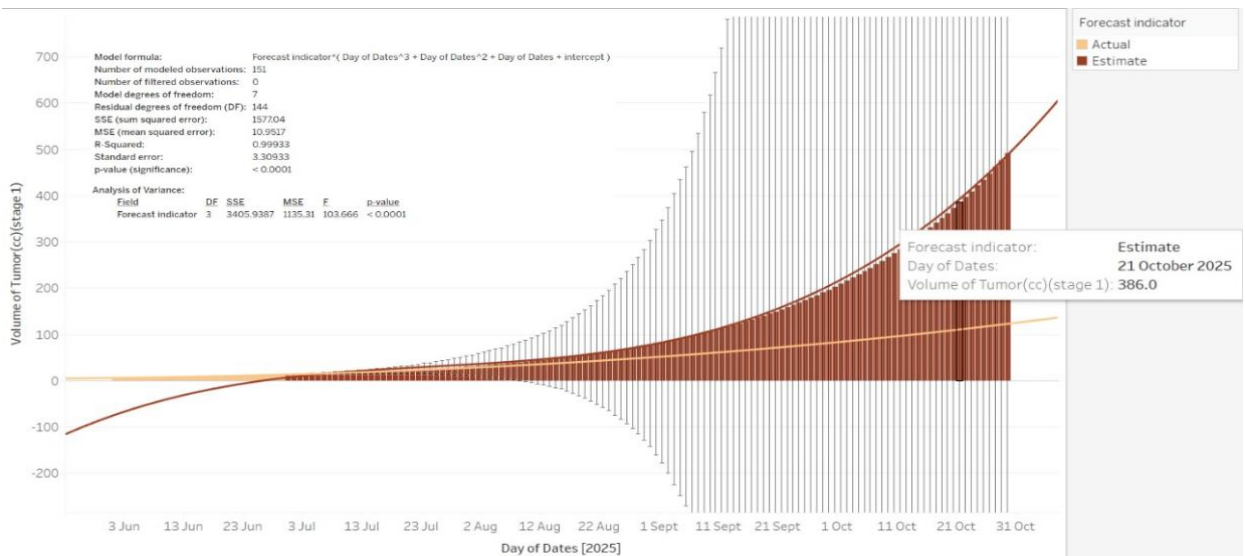

**Figure 13:** 1$^{st}$ Stage Tumor Growth Prediction

Figure 13 presents a predictive tumor kinetics graph integrated within the cognitive digital twin framework. This graph is designed to forecast the volumetric progression of a stage one brain tumor over time. The x-axis represents the timeline from early June to the end of October 2025, while the y-axis indicates the estimated tumor volume in cubic centimeters (cc), allowing for a temporal assessment of tumor behavior. The brown bars in the graph represent the model's volumetric predictions. Lighter shades indicate measurements derived from actual MRI data, while darker shades illustrate AI-based forecasted values. The chart clearly shows an upward-curving trend line, reflecting a third-degree polynomial regression model that predicts the tumor's growth over time. This trend line shows a steady increase in tumor mass. The model predicts that the tumor volume will reach 386.0 cc by October 21, 2025, as shown in the tooltip on the right. A detailed summary in the top left of the graph displays that the model is based on 151 observations, with an R-squared value of 0.9993. This indicates a near-perfect fit between predicted and actual values. The F-statistic is 103.656, and the p-value is less than 0.0001, which highlights the model's statistical importance. Additionally, error metrics like the Sum of Squared Errors (SSE), Mean Squared Error (MSE), and standard error are included, further confirming the model's predictive accuracy. This forecasting system uses historical MRI volumetric data, initial tumor volume identified through ViT++ segmentation, and real-time neurophysiological insights from EEG. The widening bars along the timeline not only show the predicted tumor volume but also indicate the growing uncertainty of long-range forecasts, warning clinicians about potential future risks. By integrating this predictive engine within the digital twin environment, clinicians can monitor tumor behavior over time, anticipate when critical thresholds might be reached, and make timely, personalized therapeutic decisions

## G) 4$^{TH}$ STAGE TUMOR PROGRESSION FORECASTING

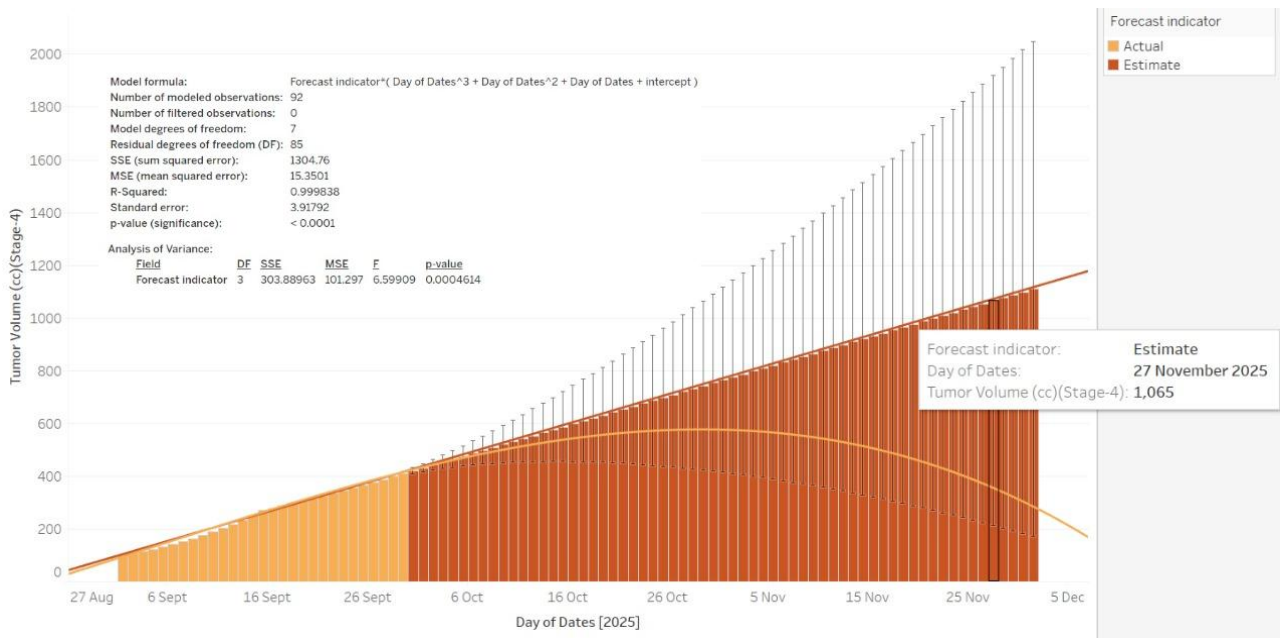

**Figure 14:** 4$^{th}$ Stage Tumor Growth Prediction

In this figure, the smooth orange curve represents the trend line. It initially rises sharply from early September, showing a rapid increase in tumor growth. This upward trend continues steadily, reaching a peak estimated volume of 1,065 cc on November 27, 2025, as shown in the tooltip. After this peak, the trend line starts to decline, indicating a slowdown in growth. This may reflect the effects of planned therapies or the limits of biological growth. The parabolic shape of the trend line captures the non-linear dynamics of tumor expansion, marked by initial rapid growth followed by periods of leveling off or decline. In the top-left corner, statistical diagnostics confirm the strength of the model based on 92 observations. An R² value of 0.9983 and a very low p-value (0.00004614) show strong predictive accuracy. Other metrics, like the F-statistic (101.297), SSE, MSE, and standard error support the model's reliability. Additionally, the widening vertical error bars along the predicted area indicate growing uncertainty in future forecasts, which is common in time-based predictions.

## VII) CONCLUSION

This research presents a transformative advancement toward next-generation neuro-oncological monitoring by proposing a real-time, scalable, and cognitively intelligent digital twin framework. By seamlessly integrating real-time EEG acquisition from a wearable skullcap, enhanced MRI analysis using an improved Vision Transformer (ViT++), bidirectional neural state classification, explainable AI through Grad-CAM, and immersive 3D brain modeling via three.js, the proposed system addresses critical gaps in current diagnostic solutions—specifically the lack of real-time adaptability, poor fusion of structural and functional modalities, and limited clinical interpretability.The system's multimodal architecture, supported by edge and fog computing layers, facilitates real-time preprocessing, risk-based filtering, and secure, low-latency data transmission. It also introduces clinically relevant forecasting capabilities, including tumor growth kinetics estimation and anomaly detection. Furthermore, the integration of explainable decision-making pipelines and intuitive 3D visualization enhances transparency and usability, enabling clinicians to interact with the digital twin environment in a meaningful, informed manner.Collectively, this work lays a strong foundation for the development of cognitively aware digital twin systems in neuro-oncology—offering the potential to reshape the diagnosis, monitoring, and management of brain disorders in modern healthcare ecosystems.

### *FUTURE WORKS*

Future works should focus on improving the capabilities of the digital digital twin by incorporating simultaneous multi-patient analysis through distributed twin orchestration and cloud-native management systems. Security and privacy of the system can be enhanced through federated learning techniques in collaborative medical environments allowing secure, decentralized model updates across institutions without direct data sharing. The EEG wearable skull cap can be upgraded by integrating vaious multiple biosensors which can aid in monitoring oxygen saturation and flow of blood inside the brain.. The accuracy of the tumor kinetics engine can be further enhanced through granular non linear modelling to capture complex tumor growth dynamics.

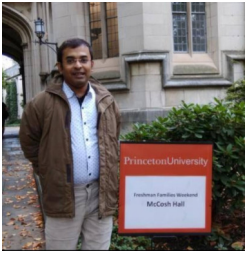

**DR. HIMADRI NATH SAHA** (Senior Member, IEEE) is a veteran researcher in the field of artificial intelligence and internet of things (IoT). He has published more than 120 research papers in different reputed journals and conferences. He also chaired many reputed IEEE conferences in the USA and CANADA and edited many IEEE conference proceedings. Dr. Saha also served as an executive member of the IEEE Kolkata, India section. He has received a good number of awards during his career, and some of them include the "Gold Faculty Award" from Infosys Technology Limited in 2013, In 2011, Infosys Technology Limited awarded him the "Outstanding Contribution Award", and in 2012, conferred him the "Best Innovative Faculty" award for implementing a unique, innovative case study on operating systems. Dr. Saha was a visiting professor at the University of Connecticut, USA. He was a visiting research scientist at Technische University Ilmenau, Germany. He is currently a visiting professor at Duale Hochschule Baden-Württemberg(DHBW) Heidenheim, Germany. His research interests include Robotics, IoT, Security, UAV, ML, WSN and Algorithm.

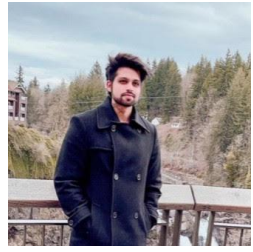

**SAPTARSHI BANERJEE** is cloud computing and artificial intelligence enthusiast dedicated to driving innovation in technology. He has completed his master's in computer science at Illinois Tech, Chicago, where his journey was marked by winning six major national hackathons. These achievements include first places at Hackillinois 2018 (University of Illinois at Urbana-Champaign), Cal Hacks 4.0 (University of California, Berkeley), MhacksX (University of Michigan Ann Arbor), Hackillinois 2017, Archhacks 2017 (Washington University in St. Louis), and Spartahacks (Michigan State University). Through these experiences, he've developed a unique perspective on crafting solutions that blend creativity, resilience, and cutting-edge tech to solve real-world challenges.

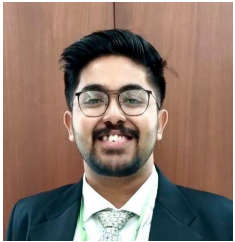

**UTSHO BANERJEE** is an undergraduate researcher pursuing his bachelor's in Computer Science and Engineering at the Institute of Engineering and Management, Kolkata. He is a passionate technophile with a strong foundation in IoT, Machine Learning, and Data Analytics. His work in AI & ML in Agritech includes image-based plant phenotyping using data-driven methods and IoT-based solutions for precision agriculture, where he devised accurate, efficient solutions that significantly improved agricultural monitoring and decision-making. Utsho has also contributed to projects in AI for healthcare, focusing on real-world applications that enhance diagnostic accuracy and system efficiency. Experienced with platforms like Tableau, and Power BI, he aims to build impactful solutions through collaboration and innovation.

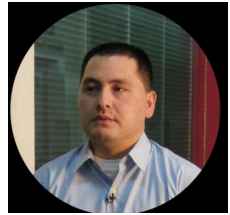

**JON TURDIEV** is a technology leader Specializing in cybersecurity, Internet of Things, and cloud computing with over two decades of industry experience. He holds an M.S. in Computer Science from San Francisco State University, where he researched IoT device security, and a B.S. in Business Administration from San Jose State University, where he graduated Magna Cum Laude. As founder and CTO of Zehntec, he architected secure solutions for healthcare, industrial automation, and transportation across North America, Europe, and Asia, with particular expertise in regulatory compliance. Previously, at Remedi Technology and JAOtech, he designed innovative medical computing systems that improved patient care in over 200 hospitals worldwide. Jon has authored multiple technical publications on cybersecurity and cloud architectures and presented at international industry conferences.

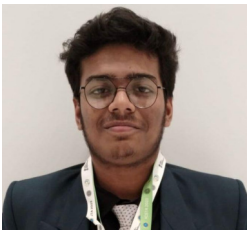

**RAJARSHI KARMAKAR** is an undergraduate researcher pursuing his bachelor's in Computer Science and Engineering at the Institute of Engineering and Management, Kolkata. He tech-driven problem solver passionate about leveraging robotics, machine learning, and data analytics to address real-world challenges in healthcare, agriculture, and emergency response. He developed an Urban Firefighting Drone featuring real-time monitoring, autonomous navigation, utilizing Coandă-effect propulsion. His work in plant phenotyping includes forecasting traits using time-series and convex hull analysis, while his medical imaging research spans ML-based classification of brain tumors, kidney conditions, and ovarian cancer. Rajarshi thrives at the intersection of data, innovation, and societal impact.